\documentclass[journal,twoside,web]{ieeecolor}
\usepackage{generic}
\usepackage{amsmath,amssymb,amsfonts}
\usepackage{graphicx}
\usepackage{textcomp}
\usepackage{amsmath,amsfonts}
\usepackage{algorithmic}
\usepackage{array}
\usepackage{pgfplots}
\usepackage{siunitx}
\usepackage{algorithmic}
\usepackage[linesnumbered,ruled,vlined]{algorithm2e}
\definecolor{r4}{HTML}{CE2029}
\definecolor{s3}{HTML}{8DB600}
\definecolor{myhund}{HTML}{BE0032}
\definecolor{myfifty}{HTML}{FF003F}
\definecolor{mytwenty}{HTML}{8B008B}
\definecolor{myzero}{HTML}{FF007F}
\definecolor{pssfhun}{HTML}{0000FF}
\definecolor{pssffif}{HTML}{333399}
\definecolor{pssftwen}{HTML}{8A2BE2}
\definecolor{pssfzero}{HTML}{6699CC}
\definecolor{r4}{HTML}{CE2029}
\definecolor{s3}{HTML}{8DB600}
\definecolor{ff}{HTML}{D2691E}
\definecolor{bf}{HTML}{FFBF00}
\definecolor{rf}{HTML}{967117}
\usepackage[T1]{fontenc}
\usepackage{pgfplots}
\usepackage{tikz}
\usepackage{tikz}
\usepackage{verbatim}
\usetikzlibrary{patterns}
\usepackage{subfigure}
\usepackage{multirow}
\usepackage{pgfplots}
\usepackage{textcomp}
\usepackage{graphicx}
\usepackage{dsfont}
\newtheorem{definition}{Definition}

\usepackage{tikz}
\usepackage{eso-pic}

\AddToShipoutPictureFG{
	\begin{tikzpicture}[remember picture,overlay]
		\node[anchor=south, yshift=8pt] at (current page.south) {
			\parbox{0.95\textwidth}{
				\centering \scriptsize \copyright~2026 IEEE. This article has been accepted in IEEE Transactions on Industrial Informatics Journal © 2026 IEEE. Personal use of this material is permitted. Permission from
				IEEE must be obtained for all other uses, in any current or future media, including reprinting/republishing this material for advertising or promotional purposes, creating new collective works, for resale or redistribution to servers or lists, or reuse of any copyrighted component of this work in other works. This work is freely available for survey and citation.
			}
		};
	\end{tikzpicture}
}

\begin{document}
\title{Multi-Factor Trust-Driven Secure Communication Model for Cloud-Based Digital Twins}
\author{Deepika Saxena, \IEEEmembership{ Senior Member, IEEE} and Ashutosh Kumar Singh, \IEEEmembership{Senior Member, IEEE}
\thanks{Deepika Saxena is with the Division of Information Systems, University of Aizu, Japan and also with Department of Computer Science, Vizja University, 01-043 Warsaw, Poland.  (Email: deepika@u-aizu.ac.jp, 13deepikasaxena@gmail.com).} 
\thanks{Ashutosh Kumar Singh is with the Department of Computer Science and Engineering, Indian Institute of Information Technology Bhopal, Bhopal 462003, India, and also with Department of Computer Science, the Vizja University, 01-043 Warsaw, Poland. (E-mail: ashutosh@iiitbhopal.ac.in).}}

\maketitle

\begin{abstract}
Cloud-based Digital Twin (DT) platforms enable real-time monitoring, simulation, and collaborative decision-making across distributed clients. However, ensuring secure and trustworthy communication remains a critical challenge due to heterogeneous client behavior, resource contention, and evolving adversarial threats. This paper proposes the \textit{Multi-Factor Trust-Driven Secure Communication} (MT-SeCom) framework to enforce resilient and intelligent collaboration in DT-enabled cloud environments. MT-SeCom operates through four coordinated phases: (i) \textit{Multi-Factor Trust Monitoring}, capturing temporal, contextual, and federated trust signals; (ii) \textit{Adaptive Trust Evaluation}, adjusting trust weights based on network dynamics and threat intensity; (iii) \textit{Transformer-Based Trusted Client Classification}, combining anomaly detection with supervised learning to accurately identify malicious or unreliable nodes; and (iv) \textit{Resilient Communication Management}, optimizing routing, isolating compromised clients, and ensuring service continuity. A real-world testbed and comprehensive experiments demonstrate that MT-SeCom significantly enhances secure communication, mitigates cascading adversarial effects, and maintains high resilience under fluctuating attack conditions. MT-SeCom achieves an average \textbf{18.7\%} improvement in threat detection accuracy and a \textbf{24.3\%} reduction in anomaly occurrences compared to existing methods, confirming its robustness, scalability, and practical suitability for heterogeneous cloud-based DT ecosystems.
\end{abstract}


\section{Introduction}
\IEEEPARstart{D}{igital} Twin (DT) technology has emerged as a cornerstone of next-generation cyber–physical systems, enabling applications such as predictive healthcare, intelligent transportation, smart energy grids, and industrial IoT ecosystems \cite{10905049}. By creating dynamic virtual replicas of physical assets, DTs enhance operational resilience, performance, and resource efficiency. However, reliance on cloud infrastructures exposes these systems to significant cybersecurity risks, as sensitive data exchanged between physical and digital counterparts can be intercepted, manipulated, or misused. Recent reports indicate that in the first half of 2025, healthcare organizations experienced 107 cyberattacks compromising over 1.6 million patient records, with Microsoft 365 as a common vulnerability \cite{woollacott2025healthcare}. The manufacturing sector accounted for 22\% of sector-attributed attacks, with ransomware incidents rising sharply \cite{bitsight2025}. Such threats highlight the critical need for robust, trust-aware, and resilient communication mechanisms. Existing solutions typically address security, anomaly detection, or trust management separately, leaving a gap in unified frameworks capable of ensuring end-to-end trustworthy communication among heterogeneous DT clients. This motivates the development of a multi-factor trust-driven approach that dynamically evaluates trust while maintaining secure, reliable, and adaptive communication in cloud-based DT networks.

\subsection{Related Work}
Recent research has explored the integration of Digital Twins (DTs) with advanced communication, security, and learning techniques to enhance performance and resilience across diverse domains. Lu \textit{et al.} \cite{lu2020communication} proposed digital twin edge networks (DITENs) that leverage federated learning and asynchronous updates to reduce communication overhead. Tang \textit{et al.} \cite{tang2025secure} developed a blockchain-based healthcare DT framework employing attribute-based encryption and missing value prediction to ensure privacy and data integrity. Soula \textit{et al.} \cite{soula2024real} introduced a DT-based trust model for IoT anomaly detection, improving scalability and energy efficiency. Other efforts have focused on blockchain-enabled resource monitoring \cite{10290991,9830718}, trust prediction in Internet of Vehicles (IoV) using Generative Adversarial Networks (GANs) \cite{liu2022blockchain}, and resilient IoT/IoV-based DT frameworks for critical cyber-physical systems \cite{saad2020implementation,9112234}. Bera \textit{et al.} \cite{bera2024digital} examined DT-driven healthcare communication security in 5G/B5G networks, addressing slice isolation and device authentication. Wang \textit{et al.} \cite{wang2023survey} provided a comprehensive survey of the Internet of Digital Twins (IoDT), detailing key architectural, communication, and security challenges.   { Mrabet et al. \cite{mrabet2025towards} have proposed multi-factor trust and resilience models that integrate anomaly detection, blockchain-based trust, privacy-preserving analytics, and secure virtualization to strengthen edge security in smart-city environments.} Collectively, these studies demonstrate that DTs can enable intelligent, secure, and efficient cyber-physical systems. However, most solutions address trust, privacy, or resilience in isolation and lack a unified framework for adaptive, end-to-end trust-aware communication across heterogeneous DT clients.

\subsection{Research Gaps and Contributions}
Despite advances in DT security and trust frameworks \cite{lu2020communication,tang2025secure,9830718,liu2022blockchain,soula2024real,saad2020implementation,9112234}, several gaps remain:  
\begin{itemize}
    \item Most trust management approaches focus on a single evaluation dimension, such as behavior monitoring, peer reputation, or cryptographic validation, which leads to an incomplete trust representation in DT communication.
    
    \item Only a few frameworks adjust trust scores in real time based on environmental changes or adversarial activities, often resulting in delayed or inaccurate trust decisions under hostile conditions.
    
    \item Trust evaluation in existing methods is often disconnected from communication resilience mechanisms, making DT systems more vulnerable to coordinated attacks.
\end{itemize}

To address these gaps, the \textit{Multi-Factor Trust-Driven Secure Communication} (MT-SeCom) model is proposed. Its main contributions include:  
\begin{itemize}
    \item The proposed framework unifies multiple trust dimensions by integrating temporal, contextual, and federated trust signals, enabling comprehensive evaluation of client reliability under dynamic and adversarial conditions.  
    \item An adaptive trust evaluation mechanism employs a Transformer-based classifier to dynamically adjust the weighting of trust factors, combining anomaly detection with supervised learning to support robust and context-aware decision-making.  
    \item The resilient communication management scheme constructs optimized, attack-aware network topologies that maximize link reliability, minimize latency, and isolate malicious clients, thereby ensuring secure and dependable communication across cloud-based DT networks.  
\end{itemize}

\noindent MT-SeCom provides a cohesive and adaptive trust-driven communication framework, unifying multi-factor trust evaluation, anomaly detection, and resilient link management to ensure end-to-end security, reliability, and performance in cloud-based DT ecosystems.

\subsection{Paper Organization}
Section \ref{Problem Formulation} presents the problem formulation including the system model, cyberthreat model, and problem statement with design goals. Section \ref{Proposed Model} details the proposed MT-SeCom framework, covering multifactor trust monitoring, adaptive trust evaluation, transformer-based client classification, and resilient communication management. Section \ref{operational design} describes the operational workflow, algorithmic summary, and complexity analysis of the proposed model. Section \ref{performance evaluations} presents the experimental setup, performance evaluation, comparative analysis, and security assessment of MT-SeCom. Finally, Section \ref{conclusion} concludes the study and outlines directions for future research.

\label{sec:introduction}
\section{Problem Formulation} \label{Problem Formulation}

This research addresses the challenge of securing collaborative cloud-based Digital Twin (DT) applications against malicious clients and unreliable behaviors while ensuring resilient communication and trustworthy execution of component tasks. The problem is formalized through: \textit{System Model} (Section~\ref{System Model}) defining participating entities and their roles; \textit{Cyberthreat Model} (Section~\ref{Cyberthreat Model}) specifying potential adversarial threats; and \textit{Problem Statement and Design Goals} (Section~\ref{Problem Statement and Design Goals}) presenting the optimization objectives and design constraints.

\subsection{System Model}\label{System Model}

As illustrated in Fig.~\ref{fig:sysmodel}, consider a system model that enables the execution of \textit{Digital Twin Applications} at the cloud platform. This system comprises  six interdependent entities: \textit{Clients} ($\mathds{C}$), \textit{Collaboration Applications} ($\mathds{A}$), \textit{Virtual Nodes} ($\mathds{VN}$), \textit{Physical Nodes} ($\mathds{PN}$), and the \textit{Cloud Platform} ($\mathds{CP}$), defined as:

\begin{figure}[!h]
    \centering
    \includegraphics[width=0.5\textwidth]{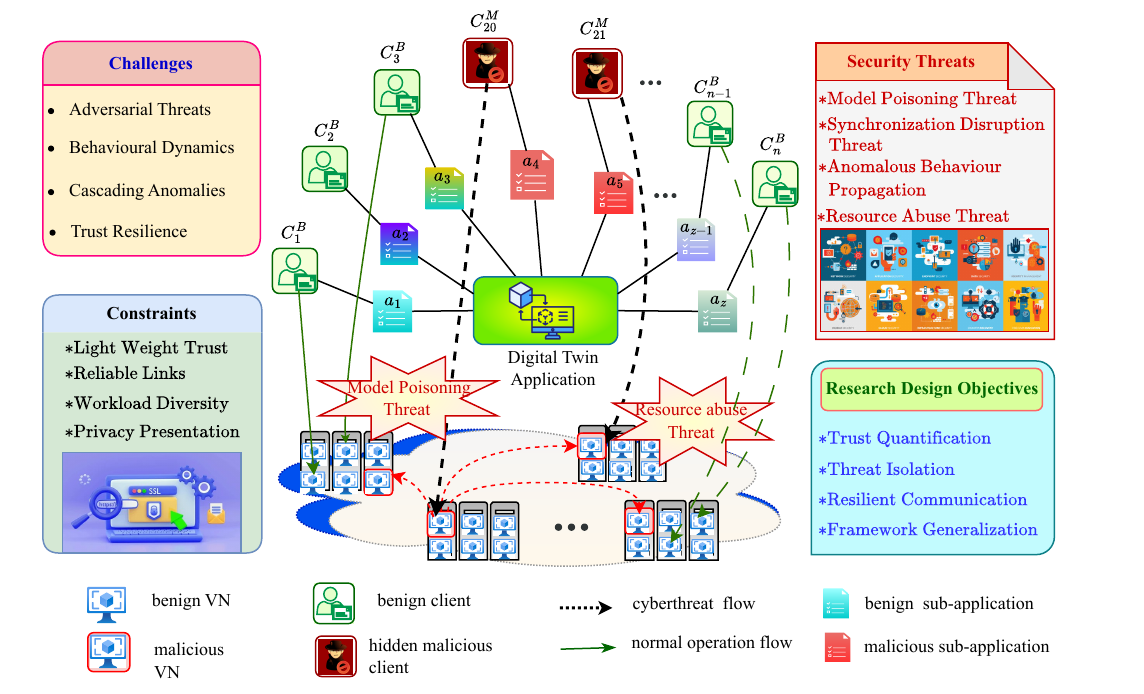}
    \caption{System model and problem illustration}
    \label{fig:sysmodel}
\end{figure}

\begin{itemize}
    \item \textit{Clients ($\mathds{C}$)}: The clients represent operators executing DT applications, $\mathds{C} = \{C_1, \ldots, C_n\}$. There are two types:
    \begin{itemize}
        \item \textit{Benign clients} ($C^B$): Legitimate participants supporting collaboration.  
        \item \textit{Malicious clients} ($C^M$): Insider adversaries that mimic benign behavior to inject malicious workloads or compromise results.  
    \end{itemize}
    
    \item \textit{Collaboration DT Applications ($\mathds{A}$)}: Each DT application comprises $z$ sub-applications, $\mathds{A} = \{A_1, \ldots, A_z\}$, executed by specific clients. Sub-applications are classified as benign or malicious based on their behavior.  

    \item \textit{Virtual Nodes ($\mathds{VN}$)}: The logical execution environments hosting sub-applications operate as virtual nodes, enabling scalability, isolation, and controlled resource allocation.  They can be benign (executing legitimate workloads) or malicious (hosting adversarial sub-applications).  

    \item \textit{Physical Nodes ($\mathds{PN}$).} 
These are physical servers that consolidate multiple virtual nodes while offering computation, storage, and networking resources required to execute Digital Twin (DT) workloads.

    \item \textit{Cloud Platform ($\mathds{CP}$)}: This platform orchestrates virtual and physical resources, ensuring elastic, resilient, and isolated execution. It supports both normal operations and propagation of cyberthreat flows.
\end{itemize}


\subsection{Cyberthreat Model}\label{Cyberthreat Model}

Consider $n$ clients $\{C_1,\ldots,C_n\}$ executing sub-applications $\{A_1,\ldots,A_z\}$ on $Q$ virtual nodes $\{VN_1,\ldots,VN_Q\}$ hosted on $P$ physical nodes $\{PN_1,\ldots,PN_P\}$, as shown in Fig.~\ref{fig:sysmodel}. A subset of hidden adversaries, denoted $C^{M}$, may control one or more malicious virtual nodes $\{VN^M_j\}$ with the objective of compromising benign components. The execution of sub-application $A_k$ on $VN_j$ running on $PN_i$ is represented by $\Omega_{ijk}: PN_i \times VN_j \times A_k$ and models a workload instance with observable timing, resource, and communication characteristics. Unless stated otherwise, threat indicators lie in $[0,1]$, with larger values indicating higher impact. This cyberthreat model comprises the following threats, and Table~\ref{Notation used in the cyberthreat model} summarizes the notation used throughout the model.

{ 
\begin{table}[!htbp] 
\caption{ {Notation used in the cyberthreat model}}
\centering
\label{Notation used in the cyberthreat model}
\resizebox{0.48\textwidth}{!}{
\begin{tabular}{lll} 
\hline
Symbol & Meaning & Range / Units \\
\hline
$PN_i$ & physical node identifier or host & integer index \\
$VN_j$ & virtual node running on $PN_i$ & integer index \\
$A_k$ & sub-application or task & integer index \\
$\Omega_{ijk}$ & execution instance of $(i,j,k)$ & dimensionless \\
$\Xi^{x}$ & threat impact score for threat $x$ & $[0,1]$ \\
$\mathds{V}^{x}_{ijk}$ & vulnerability factor for threat $x$ & $[0,1]$ \\
$\Delta t_{ijk}$ & timing deviation from nominal execution & seconds \\
$R_{ijk}$ & resource contention level & normalized, $[0,1]$ \\
$T_x^{\mathrm{federated}}(t)$ & federated trust of client $x$ at time $t$ & $[0,1]$ \\
$\Gamma(t)$ & aggregation weighting state & $[0,1]$ \\
$\Phi(\cdot)$ & aggregation function producing manipulation score & $[0,1]$ \\
\hline
\end{tabular}}
\end{table}}

\begin{definition}[Model Poisoning Threat]
Model poisoning occurs when a malicious virtual node injects corrupted parameters or manipulated gradients into collaborative learning or control updates. A malicious node $VN_{\hat{j}}$ targets sub-task $A_{\hat{k}}$ and contaminates the execution of $A_k$ running on $VN_j$ at $PN_i$. The resulting impact is represented in Eq. \eqref{e1}: 
\begin{equation}
\label{e1}
\Xi^{MP}_{i\hat{j}\hat{k} \rightarrow ijk}
=
\Omega_{ijk}
\cdot
\Omega_{i\hat{j}\hat{k}}
\cdot
PN_i
\cdot
\mathds{V}^{MP}_{ijk},
\end{equation}
where $\mathds{V}^{MP}_{ijk}\in[0,1]$ quantifies the susceptibility of the execution pipeline to poisoning. Larger values indicate weak authentication, insecure update validation, or insufficient isolation.
\end{definition}

\begin{definition}[Synchronization Disruption Threat]
Synchronization disruption captures situations where malicious virtual nodes intentionally delay, reorder, or accelerate task updates, which creates temporal inconsistencies among dependent processes. The threat is modeled in Eq. \eqref{e2}: 
\begin{equation}
\label{e2}
\Xi^{SD}_{i\hat{j}\hat{k} \rightarrow ijk}
=
\Omega_{ijk}
\cdot
\Omega_{i\hat{j}\hat{k}}
\cdot
\Delta t_{ijk}
\cdot
\mathds{V}^{SD}_{ijk},
\end{equation}
where $\Delta t_{ijk}$ measures the deviation from nominal execution time (in seconds), and $\mathds{V}^{SD}_{ijk}\in[0,1]$ expresses the degree to which such deviations can destabilize coupled DT processes.
\end{definition}

\begin{definition}[Resource Abuse Threat]
Resource abuse refers to adversarial behavior that results in abnormal consumption of CPU, memory, or network bandwidth, degrading normal system performance. The effect is captured in Eq. \eqref{e3}: 
\begin{equation}
\label{e3}
\Xi^{RA}_{i\hat{j}\hat{k} \rightarrow ijk}
=
\Omega_{ijk}
\cdot
\Omega_{i\hat{j}\hat{k}}
\cdot
R_{ijk}
\cdot
\mathds{V}^{RA}_{ijk},
\end{equation}
where $R_{ijk}\in[0,1]$ measures normalized contention level and $\mathds{V}^{RA}_{ijk}\in[0,1]$ reflects weaknesses in resource isolation that allow interference across co-located tasks.
\end{definition}

\begin{definition}[Anomalous Behavior Propagation]
Anomalous behavior propagation describes cascading effects in which local compromises produce secondary failures in dependent applications or nodes. The propagation is represented in Eq. \eqref{e4}: 
\begin{equation}
\label{e4}
\Xi^{ABP}_{i\hat{j}\hat{k}\rightarrow i^* j^* k^*}
=
\sum_{x\in\{MP,SD,RA\}}
\Xi^{x}_{i\hat{j}\hat{k}\rightarrow ijk}
\cdot
\Omega_{i^* j^* k^*}
\cdot
PN_{i^*}
\cdot
\mathds{V}_{i^* j^* k^*},
\end{equation}
where $\mathds{V}_{i^* j^* k^*}\in[0,1]$ determines how easily local anomalies propagate to downstream tasks.
\end{definition}
{ 
\begin{definition}[APT and Collusion-based Trust Manipulation]
Advanced persistent adversaries gradually accumulate high trust and later exploit it during critical operations. A malicious client $C_{\hat{i}}$, collaborating with a colluding set $\mathcal{C}$, manipulates federated trust aggregation as:
\begin{equation}
\label{eq:apt_collusion}
\Xi^{APT}_{\hat{i},\mathcal{C}}(t)
=
\Phi\!\left(
T_{\hat{i}}^{\mathrm{federated}}(t),
\{T_{c}^{\mathrm{federated}}(t)\}_{c\in\mathcal{C}},
\Gamma(t)
\right),
\end{equation}
 {where $\Phi(\cdot)$ denotes a robust operator (trimmed-mean or median). 
The trimmed-mean removes the highest and lowest $\delta\%$ values before averaging, while the median suppresses outliers. Resilience holds if the colluding fraction is below $\delta$ (or below $50\%$ for median aggregation). Here, $T_{x}^{\mathrm{federated}}(t)\!\in\![0,1]$ is the trust of client $x$, $\mathcal{C}$ is the colluding set, $\Gamma(t)\!\in\![0,1]$ controls weights, and $\Xi^{APT}_{\hat{i},\mathcal{C}}(t)\!\in\![0,1]$ denotes the manipulation magnitude.}
\end{definition}

\subsection{Problem Statement with Design Goals}\label{Problem Statement and Design Goals}

The objective is to quantify multi-factor trust for each client and sub-application, composed of temporal ($T_i^{\mathrm{temporal}}$), contextual ($T_i^{\mathrm{context}}$), and federated ($T_i^{\mathrm{federated}}$) trust, while minimizing untrusted entities and ensuring secure, resilient communication. This is formulated as follows in Eq. \eqref{e5}:

\begin{equation} \label{e5}
\begin{aligned}
& \underset{{C}_{\mathrm{trusted}}, {L}_{\mathrm{comm}}}{\text{MAX}}
& & \sum_{i \in \mathds{C}}\big( \alpha T_i^{\mathrm{temporal}} + \beta T_i^{\mathrm{context}} + \gamma T_i^{\mathrm{federated}} \big) \\
& \text{subject to} 
& & T_i^{\mathrm{temporal}}, T_i^{\mathrm{context}}, T_i^{\mathrm{federated}} \in [0,1], \quad \forall i \in \mathds{C} \\
& & & {L}_{\mathrm{comm}}(i,j) \ge L_{\min}, \quad \forall i,j \in {C}_{\mathrm{trusted}} \\
& & & {L}_{\mathrm{comm}}(i,j) \le L_{\max}^{\mathrm{latency}}, \quad \forall i,j \in {C}_{\mathrm{trusted}} \\
& & & {C}_{\mathrm{trusted}} \subseteq \mathds{C}, \quad {L}_{\mathrm{comm}} \subseteq \mathds{C} \times \mathds{C}.
\end{aligned}
\end{equation}

Here, ${C}_{\mathrm{trusted}}$ represents the set of trusted clients, ${L}_{\mathrm{comm}}$ represents communication links among them, and $\alpha, \beta, \gamma$ weight the trust dimensions. Communication links satisfy reliability and latency thresholds $L_{\min}$ and $L_{\max}^{\mathrm{latency}}$, while privacy and workload constraints ensure secure, heterogeneous DT execution.

\subsubsection*{Design Constraints}
The framework is required to operate under practical DT communication constraints. Trust evaluation must be lightweight to support real-time decisions, while communication paths must satisfy strict latency and reliability bounds. The design should accommodate heterogeneous workloads and diverse DT applications without degrading performance, and must preserve the privacy of benign clients throughout trust computation and communication.

\subsubsection*{Design Goals}
\begin{itemize}
    \item To enable accurate multi-dimensional trust evaluation for clients and tasks in DT environments.
    \item To proactively detect and isolate malicious nodes to prevent system disruption or task compromise.
    \item To ensure resilient, efficient communication among trusted entities while remaining scalable and generalizable across cloud-based DT ecosystems.
\end{itemize}

\section{Proposed  Model} \label{Proposed Model}

This section presents the \textit{Multi-Factor Trust-Driven Secure Communication} (MT-SeCom) model for secure, resilient, and optimized execution of cloud-based DT applications. MT-SeCom jointly evaluates trust, classifies clients, and adapts routing so that malicious or unstable clients are isolated while preserving performance. As illustrated in Fig. \ref{fig:proposed}, MT-SeCom enables trustworthy collaboration by operating in four phases: (\textit{i}) Multi-factor Trust Monitoring, (\textit{ii}) Adaptive Trust Evaluation, (\textit{iii}) Trusted Client Classification, and (\textit{iv}) Resilient Communication Management.

\begin{figure*}
    \centering
    \includegraphics[width=0.9\linewidth]{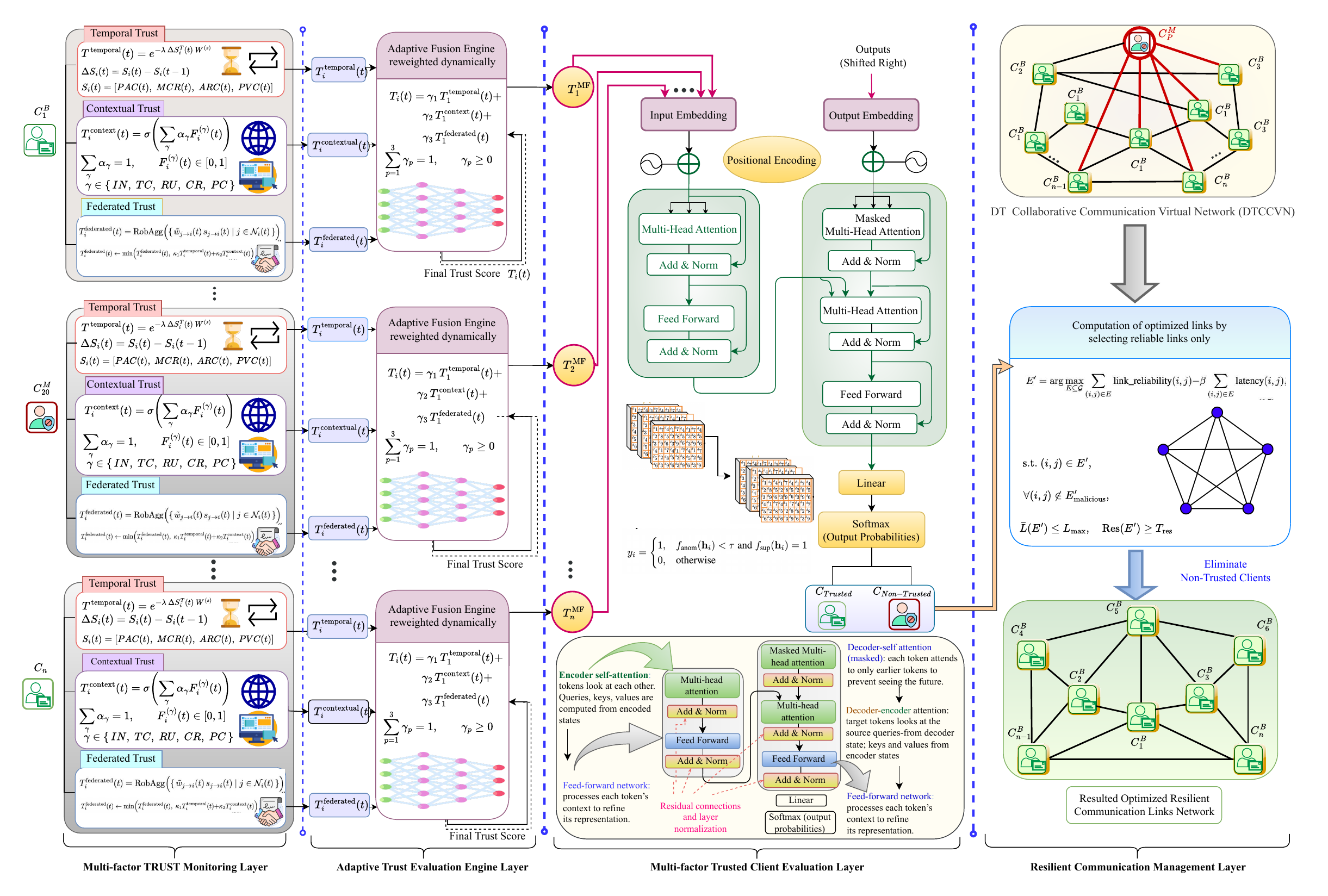}
    \caption{Proposed Model}
    \label{fig:proposed}
\end{figure*}

\subsection{Trust Monitoring Layer}

Let $n$ clients $\{C_1, ..., C_n\}$ execute sub-applications: $\{A_1, ..., A_n\}$ on $Q$ virtual nodes $\{VN_1, ..., VN_Q\}$ hosted on $P$ physical nodes $\{PN_1, ..., PN_P\}$. For each client $C_i$, multi-factor trust signals are partitioned  into three complementary components including \textit{temporal trust}, \textit{contextual trust}, and \textit{federated trust} defined as follows:
\begin{definition}[Temporal Trust]
This metric captures how a client’s performance evolves over time, penalizing negative drifts. Let $\mathbf{S}_i(t)\in\mathbb{R}^m$ denote the $m$-dimensional vector of normalized performance signals for client $i$ at time $t$, including performance accuracy ($PAC$), mission completion rate ($MCR$), average resource consumption ($ARC$), and previous violation counts ($PVC$). The temporal trust ($T_i^{\mathrm{temporal}}$) of client $i$ is defined using Eq. \eqref{eq:temporal_trust}:
\begin{equation}
    \label{eq:temporal_trust}
    T_i^{\mathrm{temporal}}(t) \;=\;
    \exp\!\Big(-\lambda \,\Delta\mathbf{S}_i^\top(t)\mathbf{w}^{(S)}\Big),
\end{equation}
with $\Delta\mathbf{S}_i(t)=\mathbf{S}_i(t)-\mathbf{S}_i(t-1)$. Here, $\lambda>0$ is a decay factor, and $\mathbf{w}^{(S)}$ is a non-negative weight vector satisfying $\mathbf{1}^\top\mathbf{w}^{(S)}=1$, reflecting the relative importance of each signal. A typical instantiation is $\mathbf{S}_i(t) =
\big[
PAC_i(t),\;
MCR_i(t),\;
ARC_i(t),\;
PVC_i(t)
\big]^\top,$ where, all elements are normalized to $[0,1]$. For negative indicators (e.g., $PVC$), normalization is inverted so that higher values consistently indicate better performance.
\end{definition}

\begin{definition}[Contextual Trust]
Contextual trust quantifies how reliable a client $i$ is under the prevailing operating
conditions. It aggregates multiple behavioral indicators capturing stability, correctness,
and communication quality. The contextual trust score at time $t$ is defined as  {Eq. \eqref{eq:contextual_trust}}:
\begin{equation}
\label{eq:contextual_trust}
T_i^{\mathrm{context}}(t)=\sigma\!\left(\sum_{r}\alpha_r F_i^{(r)}(t)\right),
\end{equation}
where $F_i^{(r)}(t)\in[0,1]$ is the normalized contribution of factor $r$,
$\alpha_r\!\ge\!0$ is its weight, and $\sum_r \alpha_r=1$.
The factors considered are $r\in$\{\textit{Output Integrity}, \textit{Timeliness},
\textit{Resource Usage}, \textit{Reliability}, \textit{Privacy Compliance}\}.
The representative formulations are:
\begin{align*}
F_i^{(\text{Output Integrity})}(t) &= 1-\mathrm{ErrRate}_i(t), \\[4pt]
F_i^{(\text{Timeliness})}(t) &= e^{-\beta \Delta t_i}, \\[4pt]
F_i^{(\text{Resource Usage})}(t) &= 1-\mathrm{clip}\!\left(\frac{\text{Usage}_i-Q_i}{Q_i},0,1\right), \\[4pt]
F_i^{(\text{Reliability})}(t) &= \frac{\text{PDR}_i}{1+\text{Jitter}_i}, \\[4pt]
F_i^{(\text{Privacy Compliance})}(t) &= 1-\text{LeakageScore}_i .
\end{align*}

Here, $\sigma(\cdot)$ is the logistic normalization that keeps
$T_i^{\mathrm{context}}(t)\!\in(0,1)$.
$\mathrm{ErrRate}_i(t)$ denotes the normalized error (e.g., accuracy loss or deviation)
in the outputs of client $i$.
$\Delta t_i$ is the reporting delay and $\beta>0$ controls its penalty.
$\text{Usage}_i$ is the observed resource consumption relative to the allocated quota $Q_i$
and is bounded by $\mathrm{clip}(\cdot)$ to $[0,1]$.
Network reliability is captured through the packet delivery ratio $\text{PDR}_i$ and
latency variability $\text{Jitter}_i$.
$\text{LeakageScore}_i$ measures the likelihood of privacy leakage.
Larger values of $T_i^{\mathrm{context}}(t)$ therefore indicate more stable,
efficient, and trustworthy behavior in the current context.
\end{definition}

{ \begin{definition}[Federated Trust]
Federated trust mitigates collusion and advanced persistent threats by
combining peer feedback with robustness and temporal decay. Let
$\mathcal{N}_i(t)$ denote the peer set of client $i$ at time $t$ as depicted in Eq. \eqref{eq:federated_robust}:
\begin{equation}
\label{eq:federated_robust}
T_i^{\mathrm{federated}}(t)
=
\mathrm{RobAgg}
\Big(
\{\, \tilde{w}_{j\to i}(t)\, s_{j\to i}(t)
\mid j \in \mathcal{N}_i(t)\,\}
\Big),
\end{equation}
where $s_{j\to i}(t)\in[0,1]$ is the feedback from peer $j$, and
$\mathrm{RobAgg}(\cdot)$ denotes a trimmed-mean or median operator that
suppresses extreme opinions and reduces collusion. The credibility weight is given by Eq. \eqref{fed_trust}:
\begin{equation}
\label{fed_trust}
\tilde{w}_{j\to i}(t)
=
\eta e^{-\rho \Delta t_j}
+
(1-\eta)
\frac{1}{1+\mathrm{Var}(T_j(1{:}t))},
\end{equation}
where $e^{-\rho \Delta t_j}$ enforces time decay so that recent behavior
contributes more strongly, and $\mathrm{Var}(T_j(1{:}t))$ penalizes peers
exhibiting unstable trust trajectories. The parameter $\eta \in [0,1]$
controls the balance between recency and stability.

To ensure that peer opinions do not overshadow direct evidence,
$T_i^{\mathrm{federated}}(t)$ is bounded as Eq. \eqref{kappa}:
\begin{equation}
\label{kappa}
T_i^{\mathrm{federated}}(t)
\leftarrow
\min\!\Big(
T_i^{\mathrm{federated}}(t),\;
\kappa_1 T_i^{\mathrm{temporal}}(t)
+
\kappa_2 T_i^{\mathrm{context}}(t)
\Big)
\end{equation}
with $\kappa_1 + \kappa_2 = 1$. This design limits gradual trust
accumulation, attenuates colluding peers, and prevents federated trust from
overriding temporal or contextual evidence.
\end{definition}

MT-SeCom updates trust incrementally using sliding windows, retaining only
recent telemetry. The per-client update cost remains $O(m)$, where $m$
denotes the number of monitored factors. This incremental design enables
continuous trust estimation with low memory and computational overhead, even
at large scale.  {To defend against collusion and gradual trust inflation, federated aggregation combines (i) robust statistics (trimmed-mean or median), (ii) credibility-aware weighting}  {with temporal decay and variance penalization, and (iii) an upper bound that constrains federated trust by temporal and contextual evidence. The aggregation remains resilient provided the fraction of colluding peers does not exceed the robustness tolerance (e.g., trimming parameter in trimmed-mean or $50\%$ in median aggregation).} As a result, malicious feedback cannot dominate aggregation unless adversarial participation exceeds the robustness tolerance threshold.

\subsection{Adaptive Multi-Factor Trust Evaluation}

The multi-factor trust score for client $C_i$ is computed dynamically as Eq. \eqref{eq:trust}, where $\gamma_p \ge 0$ and $\sum_{p=1}^3 \gamma_p = 1$.
\begin{equation} \label{eq:trust}
  T_i(t) = \gamma_1\,T_i^{\mathrm{temporal}}(t)
         + \gamma_2\,T_i^{\mathrm{context}}(t)
         + \gamma_3\,T_i^{\mathrm{federated}}(t),
\end{equation}

The trust engine continuously updates $T_i(t)$ in real time. Temporal and
contextual trust values are derived from performance and environmental
signals, while federated trust is aggregated from peer feedback. A feedback
controller adaptively adjusts $(\gamma_1,\gamma_2,\gamma_3)$: contextual
trust is emphasized under network instability, temporal trust dominates for
long-term reliability, and federated trust is prioritized in uncertain or
adversarial conditions based on observed violation rates, link instability,
and anomaly frequency.  {Because each trust component is updated locally and independently, the trust
engine aggregates compact summaries rather than full historical traces. This
reduces recomputation overhead and allows $T_i(t)$ to scale efficiently with
both the number of clients and the event rate.
}

\subsection{Trusted Client Classification}

Client trust is inferred using a Transformer-based classifier $F_\theta$ that
maps the multi-factor signal vector $\mathbf{S}_i$ to a binary decision
(Eq.~\ref{e9}):
\begin{equation} \label{e9}
    y_i = F_\theta(\mathbf{S}_i) \in \{0,1\}, \quad 
    y_i = 
    \begin{cases}
        1, & \text{if client $i$ is trusted}, \\
        0, & \text{if client $i$ is malicious}.
    \end{cases}
\end{equation}

 {
The Transformer-based classifier operates on compact multi-factor trust
features instead of raw telemetry streams. This keeps inference linear in
sequence length and easily parallelizable across clients, enabling MT-SeCom
to sustain high event rates without introducing noticeable latency in the
decision pipeline.
} The Transformer encoder models temporal and contextual dependencies in
$\mathbf{S}_i$, yielding an embedding
$\mathbf{h}_i = \text{Transformer}_\theta(\mathbf{S}_i)$.
Classification integrates supervised prediction and anomaly detection via
Eq.~\eqref{e10}:
\begin{equation} \label{e10}
    y_i = 
    \begin{cases}
        1, & f_{\text{anom}}(\mathbf{h}_i) < \tau \;\;\text{and}\;\; f_{\text{sup}}(\mathbf{h}_i) = 1, \\
        0, & \text{otherwise},
    \end{cases}
\end{equation}
where $f_{\text{anom}}(\cdot)$ denotes the anomaly score,
$f_{\text{sup}}(\cdot)$ the supervised classifier output, and $\tau$ the
detection threshold. This joint decision rule reduces false acceptance of
anomalous clients while preserving sensitivity to legitimate behavior.

 {To exploit temporal trust dynamics, the Transformer encoder applies multi-head self-attention over trust sequences. Unlike recurrent models, self-attention avoids stepwise state propagation and captures long-range dependencies without vanishing gradients. The mechanism emphasizes salient trust events (e.g., performance drifts, policy violations, reputation drops) while suppressing benign fluctuations. By jointly encoding temporal, contextual, and federated indicators, the model learns their interactions and detects patterns formed by multiple weak signals, yielding a more discriminative representation of client behavior and more reliable trust decisions in MT-SeCom.
}

\subsection{Resilient Communication Management}

The \textit{Resilient Communication Manager (RCM)} maintains secure and efficient connectivity among trusted DT clients by adapting the communication graph network  $\mathcal{G}(V,E)$, where $V$ denotes clients with $y_i=1$ and $E$ represents active links. At each decision step, the RCM derives a resilient subgraph $\mathcal{G}'=(V \setminus V_{\text{malicious}},E')$ by balancing reliability and latency while isolating links associated with suspicious nodes. The resulting link set satisfies using Eqs. \eqref{rcm1} and \eqref{rcm2}:
\begin{equation} \label{rcm1}
E' = \arg\max_{E \subseteq \mathcal{G}}
\sum_{(i,j)\in E} \text{link\_reliability}(i,j)
- \beta \sum_{(i,j)\in E} \text{latency}(i,j),
\end{equation}
subject to
\begin{equation} \label{rcm2}
(i,j)\notin E \;\forall (i,j)\in E_{\text{malicious}}, \quad
\bar{L}(E)\le L_{\max}, \quad
\text{Res}(E)\ge \tau_{\text{res}}.
\end{equation}

{
 
Because solving this optimization exactly under dynamic conditions is expensive, MT-SeCom approximates it using reinforcement learning.
The system state $s_t$ summarizes the current network context, including trust scores, link reliabilities, queue backlogs, and average path delay. An action $a_t$ modifies the active communication network  by activating, deactivating, or rerouting selected trusted links. The reward function encourages resilient yet efficient routing and is given by Eq. \eqref{rl}:
\begin{equation} \label{rl}
r_t = \alpha_1 \,\text{Res}(E_t)
- \alpha_2 \,\bar{L}(E_t)
- \alpha_3 \,\text{penalty}_{\text{mal}}(E_t),
\end{equation}
where, the resilience term reflects path survivability under failures, the latency term penalizes delay, and the final term discourages routes traversing low-trust nodes. Learning is implemented using a model-free actor–critic scheme that updates the policy from transitions $(s_t,a_t,r_t,s_{t+1})$. Training continues until the moving average of cumulative reward stabilizes within a tolerance window, indicating convergence. During operation, the policy adapts gradually as traffic conditions and trust dynamics evolve, enabling the RCM to reduce exposure to unreliable links while maintaining communication constraints.
To further reduce computational cost, the RCM operates only on the pruned network consisting of trusted clients. As a
result, routing optimization scales with the size of the active trusted
sub-network (approximately $O(|E'|\log|V|)$) rather than the entire network.
This network-pruning strategy substantially decreases routing complexity while preserving resilience guarantees.
}

\section{Operational Design and Complexity Analysis}
\label{operational design}

The operational workflow of MT-SeCom is summarized in Algorithm~1. Each
phase is designed to operate incrementally and in parallel, thereby enabling
scalable execution under high client populations and event rates.

\begin{algorithm}[!h]
\DontPrintSemicolon

\tcc{\textbf{Multi-Factor Trust Monitoring}}
\For{each client $C_i$}{
Update temporal trust $T_i^{\mathrm{temporal}}$ using Eq.~\eqref{eq:temporal_trust}\\
Update contextual trust $T_i^{\mathrm{context}}$ using Eq.~\eqref{eq:contextual_trust}\\
Update federated trust $T_i^{\mathrm{federated}}$ via Eq.~\eqref{eq:federated_robust}
}

\tcc{\textbf{Adaptive Trust Evaluation}}
\For{each client $C_i$}{
Compute overall trust:
$T_i = \gamma_1 T_i^{\mathrm{temporal}}
     + \gamma_2 T_i^{\mathrm{context}}
     + \gamma_3 T_i^{\mathrm{federated}}$\\
Adapt $(\gamma_1,\gamma_2,\gamma_3)$ based on violation rate, instability, and anomaly trends
}

\tcc{\textbf{Trusted Client Classification}}
\For{each client $C_i$}{
Compute embedding $\mathbf{h}_i=\text{Transformer}_\theta(\mathbf{S}_i)$\\
Classify using Eq.~\eqref{e10}
}

\tcc{\textbf{Resilient Communication Management}}
Construct trusted set $V_{\text{trusted}}=\{C_i\mid y_i=1\}$\\
Prune malicious nodes and links to obtain sub-network $\mathcal{G}'$\\
Optimize $E'$ under resilience/latency constraints using RL policy\\
Deploy updated topology $\mathcal{G}'=(V_{\text{trusted}},E')$

\caption{Operational Summary of MT-SeCom}
\end{algorithm}

\subsection*{Complexity Analysis}

MT-SeCom is designed for scalable operation through incremental updates, compact feature processing, and trusted network optimization. Trust metrics are computed locally using sliding windows, requiring only recent telemetry and yielding $\mathcal{O}(n\cdot m)$ time and $\mathcal{O}(n)$ space, where $n$ is the number of clients and $m$ the monitored signals. Adaptive weighting introduces only $\mathcal{O}(n)$ overhead. The Transformer-based classifier operates on short trust sequences, resulting in $\mathcal{O}(n\cdot L)$ inference cost with $\mathcal{O}(n\cdot d)$ memory. Communication optimization is performed only over the pruned trusted subgraph $\mathcal{G}'=(V',E')$, with routing and refinement scaling as $\mathcal{O}(|E'|\log|V'|)$ and reinforcement-learning updates as $\mathcal{O}(|E'|\cdot T_{\text{RL}})$. Overall, MT-SeCom achieves low computational and memory overhead while supporting large client populations and high event rates.

\section{Performance Evaluation} \label{performance evaluations}

\subsection{Experimental Set-up}

The simulation experiments are conducted on a high-performance server equipped with dual Intel\textsuperscript{\textregistered} Xeon\textsuperscript{\textregistered} Silver 4114 CPUs (40 cores, 2.20\,GHz). The setup synthesizes realistic VM time-series data by concatenating multiple per-VM CSV traces (stored in \texttt{GCD\_VMs}) \cite{reiss2011google}, each containing CPU and memory utilization values normalized and windowed into sequences of length \texttt{SEQ\_LEN} for Transformer-based anomaly prediction.  {To train the Transformer classifier, ground-truth labels are constructed automatically during attack injection. Each VM sequence is labeled as \texttt{trusted} when it evolves under normal conditions, and \texttt{malicious} when the \texttt{simulate\_attack} routine injects abnormal behavior. The classifier is therefore trained in a supervised manner on binary labels \(\{0,1\}\). The dataset is split into non-overlapping training, validation, and test partitions using a 70\,:\,15\,:\,15 protocol, while ensuring that sequences originating from the same VM do not appear across different splits. The Transformer architecture (\texttt{D\_MODEL}, \texttt{NHEAD}, \texttt{NUM\_LAYERS}) is optimized using the Adam optimizer with learning rate \texttt{LR}. Binary cross-entropy is used as the classification loss, and model selection is guided by validation loss with early stopping patience of five epochs to prevent overfitting. For anomaly prediction, reconstruction error is trained using MSE loss, followed by a median + Median Absolute Deviation (MAD) thresholding rule for anomaly scoring. Thresholds are fixed across experiments for reproducibility.} Table~\ref{tab:exp-params} summarizes the experimental configuration. VM-to-PM assignment is performed via \texttt{allocate\_vms\_pm}, parameterized by \texttt{VM\_COUNT} and \texttt{PM\_COUNT}, while system metadata (for example \texttt{Model\_Drift}, \texttt{PDR}, \texttt{CPU\_cap}) is generated using a fixed seed for reproducibility.  {Attacks are injected using \texttt{simulate\_attack} with a Bernoulli distribution governed by \texttt{ATTACK\_RATE}, resulting in stochastic injection at random time steps during operation. When triggered, the injected attack persists for a configurable duration window (\texttt{ATTACK\_DURATION}), thereby modeling sustained malicious behavior rather than single isolated spikes. This design enables evaluation under both intermittent and continuous adversarial conditions.} Trust metrics (Contextual, Temporal, Federated, and Adaptive) are computed to emulate dynamic trust-aware decision-making, where VMs with \texttt{Adaptive\_Trust} $>0.5$ guide a greedy optimizer for selecting communication paths under latency and resource constraints. 

\begin{table}[!h]
  \caption{Experimental parameters and their values}
  \label{tab:exp-params}
  \centering
  \resizebox{0.48\textwidth}{!}{
  \begin{tabular}{ll}
    \hline
    \textbf{Parameter} & \textbf{Default / Range} \\
    \hline
    \texttt{VM\_COUNT} & 100-1600 VNs \\
    \texttt{PM\_COUNT} & 30-500 \\
    \texttt{SEQ\_LEN} & 10 \\
    \texttt{ATTACK\_RATE} & 0.05 \\
    RNG Seed & 42 \\
    Transformer architecture & \texttt{D\_MODEL}=64, \texttt{NHEAD}=4, \texttt{NUM\_LAYERS}=2 \\
    Training hyperparameters & \texttt{BATCH\_SIZE}=128, \texttt{EPOCHS}=5, LR=1e-3 \\
    Thresholding rule & median + 3*Median Absolute Deviation (MAD) \\
    Link selection constraints & \(\text{L\_max}=150\), \(\tau_{\text{res}}=0.1\) \\
    \hline
  \end{tabular}}
\end{table}

\subsection{MT-SeCom Results}
The performance metrics in  {Table~\ref{table:performance_mtsecom}} demonstrate the scalability and robustness of the proposed \textit{MT-SeCom} framework under increasing cloud–DT workloads. Trust evaluation accuracy remains above $93\%$ and peaks at $95.20\%$ for 800~VNs, while the mean error stays extremely low ($<7\times10^{-4}$). This accuracy–error consistency indicates the effectiveness of the adaptive multi-factor trust computation (Eq.~\eqref{eq:trust}) in modeling evolving behavioral patterns across temporal, contextual, and federated dimensions. Furthermore, the low Hamming loss ($\textit{HL} < 0.08$) confirms the reliability of the Transformer-enhanced trust classifier in detecting fine-grained deviations. From a system efficiency standpoint, \textit{MT-SeCom} maintains stable communication and resource utilization as scale increases. Power usage decreases with larger deployments, indicating reduced per-node processing overhead enabled by the lightweight trust-calculation pipeline. Resource utilization remains balanced (\(63\text{–}66\%\)), ensuring effective task–resource coordination without overloading physical infrastructure. Therefore, the framework demonstrates strong stability and efficiency even as workload scale increases by more than an order of magnitude.

  \begin{table}[!h]
 	\caption[Table caption text] {Performance metrics for MT-SeCom over increasing VNs.} 
  \centering
 	\label{table:performance_mtsecom}
 	\small
 	\resizebox{9cm}{!}{
 		
 		\begin{tabular}{|l|c|c|c|c|c|c|c|c|c|c|c|}
 			\hline
 			
 		\textit{VN}&$APN$& $T$&$Acc$& $ME \times 10^{-4}$&$PW$(W) &$RU$(\%) & $HL$  \\ \hline 
100 & 30 & 85.04 & 93.47&5.53 & 6.12E-03 & 66.05 & 0.064 \\ \hline

400 & 110 & 85.03& 94.56& 6.89 & 1.96E-04&64.07 & 0.037\\ \hline

600 & 180 & 85.07& 93.33&5.98 &2.17E-04 & 64.15& 0.027\\ \hline

800 & 210&86.04 & 95.20&4.95 &3.31E-04 &63.04 & 0.078\\ \hline

1000 & 300& 87.03&94.44 &4.05 &3.59E-04 & 63.87& 0.052\\ \hline

 {1600} &455 &87.07 & 94.36&3.89 & 3.92E-04&63.32 & 0.045\\ \hline
        
 	\end{tabular}}
     \footnotesize{\scriptsize{\textit{VN}: Virtual node; $APN$: active physical node; $Acc$: Average accuracy; $ME$: Mean error; $PW$: Power consumption; $RU$: Resource utilization; $HL$: Hamming loss }}
 \end{table}

Fig.~\ref{fig:trust results} illustrates the behavior of the multi-factor trust mechanism under increasing adversarial intensity. Across all attack levels ($5\%$--$90\%$) and deployment scales (100--{1600}~VNs), \textit{MT-SeCom} consistently maintains higher trust scores compared to baseline models. Temporal trust in Fig.~\ref{fig:trust results}(a) decreases gradually as attack intensity rises, reflecting the exponential penalization of behavioral drift in~(\ref{eq:temporal_trust}) and enabling rapid identification of compromised clients. Contextual trust in Fig.~\ref{fig:trust results}(b) remains stable despite fluctuating workloads and adversarial behavior, supported by continuous normalization of communication, resource, and privacy metrics in~ {(Eq. \ref{eq:contextual_trust})}. 
 \begin{figure*}[!h]
\centering	
\subfigure[Temporal trust]{
\includegraphics[width=0.23\linewidth, scale=8]{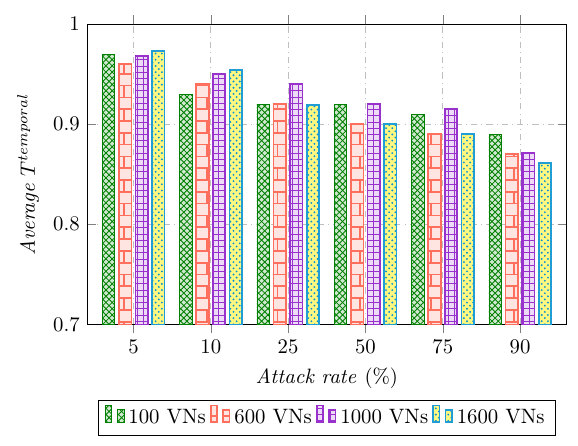}  
}
\subfigure[Contextual trust]{\includegraphics[width=0.23\linewidth, scale=8]{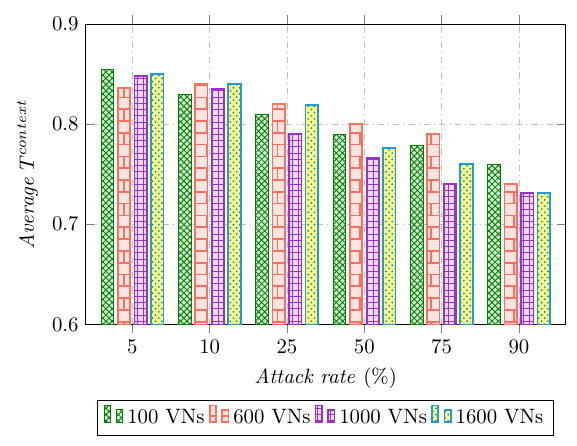}
}
\subfigure[Federated trust ]{
  \includegraphics[width=0.23\linewidth, scale=8]{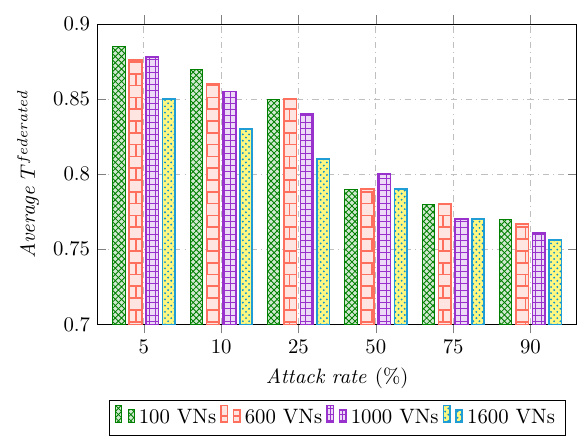}}
\subfigure[Adaptive multi-factor trust ]{
  \includegraphics[width=0.23\linewidth, scale=8]{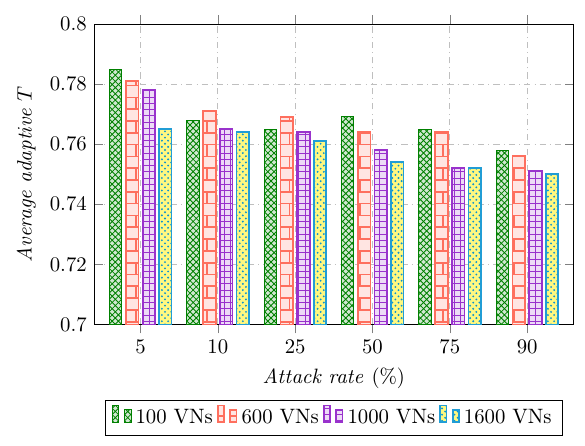}}
\caption{Multi-factor trust analysis}
\label{fig:trust results}
\end{figure*}
Meanwhile, Fig.~\ref{fig:trust results}(c) shows that federated trust mechanisms dampen malicious influence through collaborative anomaly detection and validator consensus. The fused behavior in Fig.~\ref{fig:trust results}(d) demonstrates the strength of adaptive multi-factor aggregation. Through dynamic adjustment of weights $(\gamma_1,\gamma_2,\gamma_3)$ in~(\ref{eq:trust}), the model prioritizes the most reliable trust channel as adversarial conditions shift, maintaining separation between legitimate and malicious nodes. Overall, the results confirm that resilient trust in large-scale DT environments is best achieved through adaptive, context-aware fusion rather than reliance on a single trust dimension.
\par Fig.~\ref{fig:pr-roc} summarizes the operational performance of \textit{MT-SeCom} under increasing attack intensity and system scale. In Fig.~\ref{fig:pr-roc}(a), the Transformer-based trust classifier sustains high accuracy ($>90\%$ at low attack rates and $>80\%$ even at $90\%$ attacks), indicating reliable discrimination between benign and malicious clients despite adversarial noise. Consistently, the low Hamming loss trend in Fig.~\ref{fig:pr-roc}(b) confirms minimal false approvals or rejections. Together, these metrics validate the effectiveness of the hybrid anomaly–classification strategy, where temporal–context learning supports fine-grained behavior modeling and anomaly scoring prevents trust inflation from compromised clients.
\begin{figure*}[!h]
\centering	
\subfigure[Trusted client classification accuracy]{
\includegraphics[width=0.23\linewidth, scale=8]{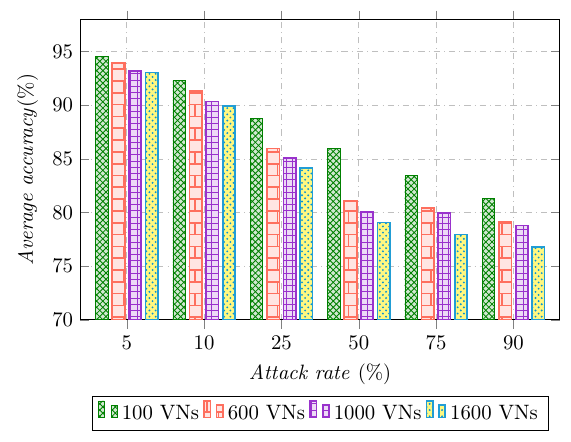}  
}
\subfigure[Hamming loss]{\includegraphics[width=0.23\linewidth, scale=8]{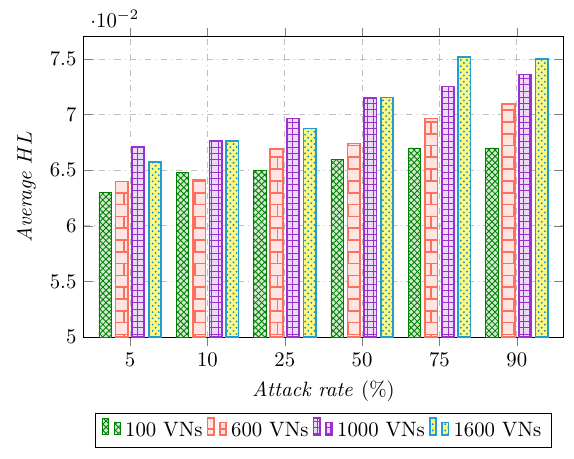}
}
\subfigure[Trust improvement]{
  \includegraphics[width=0.23\linewidth, scale=8]{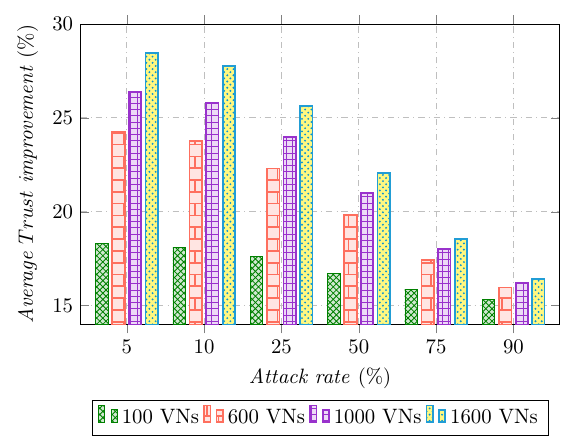}}
\subfigure[Latency reduction]{
  \includegraphics[width=0.23\linewidth, scale=8]{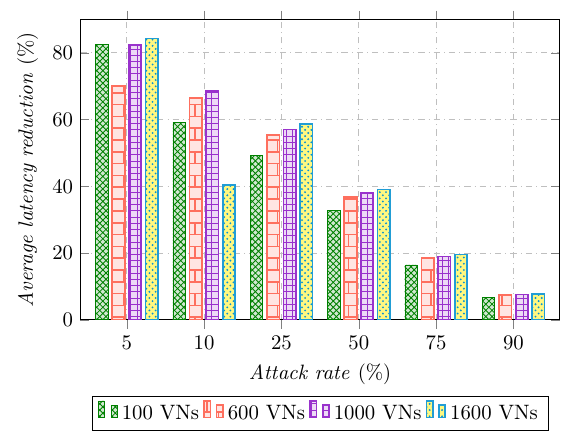}}
\caption{MT-SeCom efficiency analysis}
\label{fig:pr-roc}
\end{figure*}
The benefits of the Resilient Communication Manager (RCM) are reflected in Fig.~\ref{fig:pr-roc}(c)--(d). Trust improvement increases with attack severity as the system selectively isolates low-trust nodes and reroutes communication links through reliable peers. Simultaneously, the communication latency reduction (up to $80\%$ at low attack levels and $>40\%$ even under $90\%$ attacks) confirms that the RL-driven optimization not only preserves defense integrity but also enhances communication efficiency. Overall, the results demonstrate that \textit{MT-SeCom} maintains secure, adaptive, and scalable trust-based collaboration even under extreme adversarial stress.
\subsection{Ablation Analysis}
Table~\ref{table:performanceGCD} presents the ablation analysis of \textit{MT-SeCom} across six architectural variants to quantify the contribution of each trust component and mechanism. The full model demonstrates optimal balance with $94.23\%$ accuracy, $F_1{=}0.8231$, and a minimal mean error ($5.06\times10^{-4}$), confirming the complementary value of temporal, contextual, and federated trust integration.

  \begin{table}[!h]
 	\caption[Table caption text] {MT-SeCom Ablation Analysis.} 
  \centering
 	\label{table:performanceGCD}
 	\small
 	\resizebox{9cm}{!}{
 		
 		\begin{tabular}{|l|c|c|c|c|c|c|c|c|c|c|c|}
 			\hline
 			
 		\textbf{MT-SeCom versions} &$Acc$& $Prec.$&$Recall$& $F1$& $HL$&$ROC-AUC$&$ME$   \\ \hline 
MT-SeCom (Full) & 94.23 & 0.95004 & 0.8310&0.8231 & 0.0644& 0.9500 & 5.06E-04  \\ \hline

MT-SeCom+No $T^{temporal}$ & 92.65 & 0.95005& 0.8491& 0.8330 & 0.0732&0.9500 & 4.93E-05\\ \hline

MT-SeCom+No $T^{context}$  & 93.35 & 0.95039& 0.8432&0.8341 &0.0414 & 0.9501& 6.76E-04\\ \hline

MT-SeCom+No $T^{federated}$  & 92.04&0.95016 & 0.8562& 0.8391 &0.0262&0.9500& 4.11E-04\\ \hline

MT-SeCom+No adaptive trust & 96.33& 0.94991&0.8566&0.8041 &0.0367 & 0.9601& 3.21E-04\\ \hline

Adaptive Trust+MLP &91.83 &0.95012 & 0.8588 &0.8390 & 0.0813&0.97003 & 5.34E-03\\ \hline
        
 	\end{tabular}}
     \footnotesize{\scriptsize{\textit{VN}: Virtual node; $APN$: active physical node; $Acc$: Average accuracy; $ME$: Mean error; $PW$: Power consumption; $RU$: Resource utilization; $HL$: Hamming loss }}
 \end{table}
 Removing individual trust factors leads to measurable degradation, excluding $T^{\mathrm{temporal}}$ and $T^{\mathrm{federated}}$ lowers accuracy to $92.65\%$ and $92.04\%$, respectively, while omission of $T^{\mathrm{context}}$ reduces $F_1$-score, reflecting its role in situational adaptability. Notably, the “No adaptive trust’’ variant yields higher accuracy ($96.33\%$) but a weaker $F_1$-score ($0.8041$), demonstrating that static weighting overfits and lacks robustness under dynamic conditions. Meanwhile, replacing the Transformer with an MLP significantly degrades performance ($Acc{=}91.83\%$), highlighting the necessity of attention-driven sequence modeling. 
\subsection{Comparison}

The comparative performance analysis in Fig.~\ref{fig:comparison} demonstrates the superiority of the proposed MT-SeCom framework across multiple evaluation dimensions under varying attack intensities ranging from $5\%$ to $90\%$. As shown in Fig.~\ref{fig:comparison}(a), MT-SeCom consistently achieves the highest accuracy, maintaining performance above $85\%$ even under extreme adversarial loads (i.e., $>75\%$ attack rate). In contrast, baseline models such as PCA, Isolation Forest, and OCSVM experience a sharp degradation as attack rates increase, reflecting their limited adaptability to dynamic and coordinated adversarial behaviors. The Transformers-based trust-driven classifier in MT-SeCom enables stronger resilience by jointly modeling temporal drifts, contextual signatures, and federated trust feedback, thereby preventing false trust elevation of malicious clients.

\begin{figure*}[!h]
\centering	
\subfigure[Accuracy]{
\includegraphics[width=0.31\linewidth, scale=8]{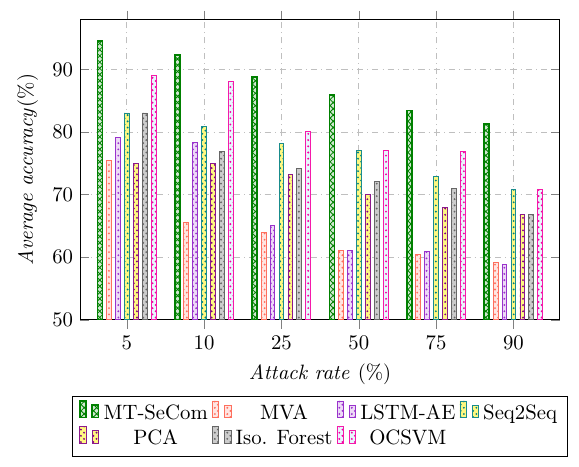}  }
\subfigure[Hamming loss]{\includegraphics[width=0.31\linewidth, scale=8]{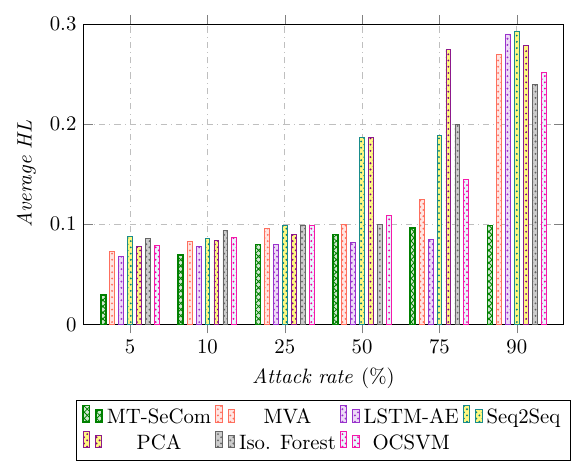}
}
\subfigure[Precision]{
  \includegraphics[width=0.33\linewidth, scale=8]{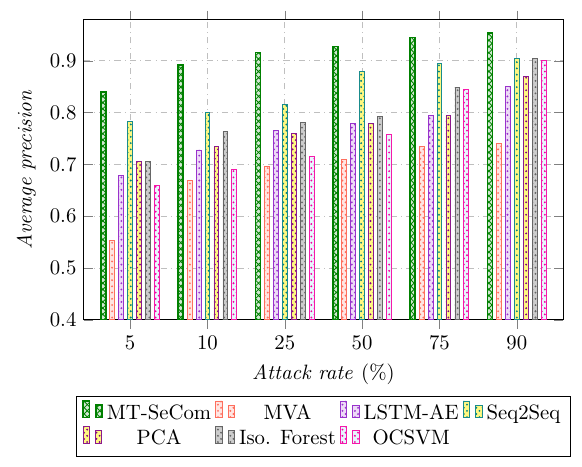}}
\subfigure[F1 Score]{
  \includegraphics[width=0.31\linewidth, scale=8]{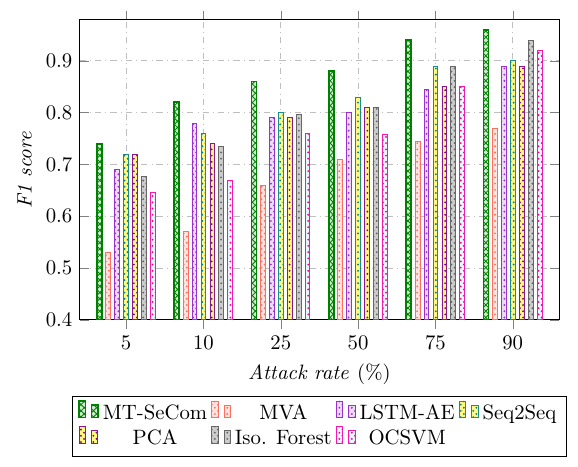}}
\subfigure[ROC-AUC]{
  \includegraphics[width=0.31\linewidth, scale=8]{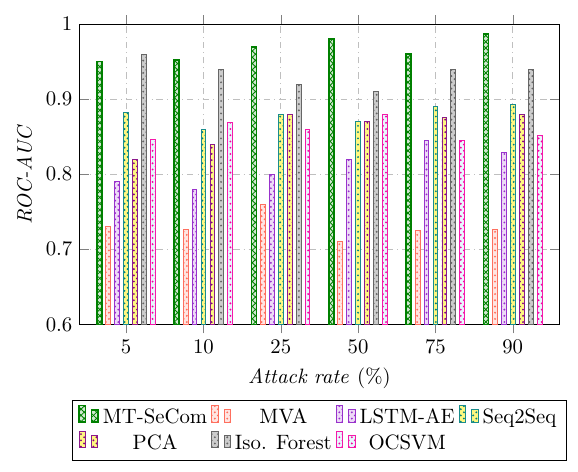}}
\subfigure[Anomaly occurrence]{
  \includegraphics[width=0.31\linewidth, scale=8]{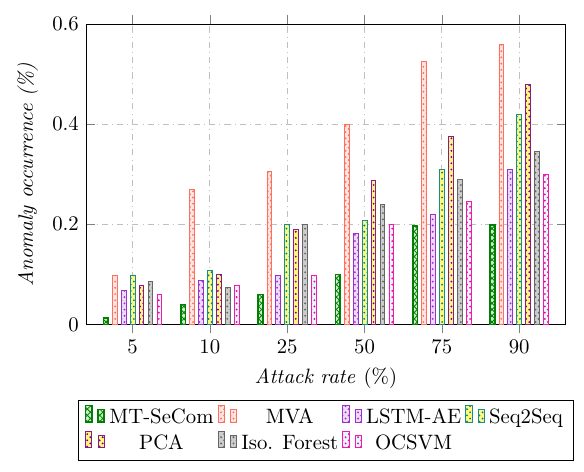}}
\caption{Clients classification analysis: MT-SeCom versus state-of-the-art methods.}
\label{fig:comparison}
\end{figure*}
Fig.~\ref{fig:comparison}(b) shows that MT-SeCom achieves the lowest Hamming loss across all attack intensities. Classical anomaly detectors and shallow learning methods perform adequately under mild adversarial conditions ($<25\%$) but degrade sharply beyond $50\%$. MT-SeCom maintains stable performance via adaptive trust–weighted inference, mitigating misclassification propagation. Precision results (Fig.~\ref{fig:comparison}(c)) remain near $0.9$ even under high adversarial noise, outperforming deep models (LSTM-AE, Seq2Seq) that lack federated trust alignment. F1-score and ROC-AUC trends (Figs.~\ref{fig:comparison}(d)--\ref{fig:comparison}(e)) show balanced detection of malicious and legitimate behavior, with ROC-AUC consistently above $0.9$, thanks to dynamic trust parameter adjustment $(\gamma_1,\gamma_2,\gamma_3)$. Fig.~\ref{fig:comparison}(f) confirms that MT-SeCom minimizes anomaly occurrence under extreme attacks ($>75\%$), suppressing cascading adversarial effects, whereas reconstruction-based baselines overflag anomalies. 

\section{Conclusion and Future Work} \label{conclusion}

This paper proposed the \textit{Multi-Factor Trust-Driven Secure Communication} (MT-SeCom) model for secure and resilient collaboration in cloud-based Digital Twin (DT) networks. By combining temporal, contextual, and federated trust with a Transformer-based classifier, MT-SeCom effectively identifies trusted clients and isolates malicious nodes, while the Resilient Communication Manager ensures reliable, low-latency communication. Analytical evaluation demonstrates its adaptability to dynamic network conditions and adversarial threats, offering a scalable framework for heterogeneous DT environments.

 {Future work focuses on richer attack models such as bursty, periodic, adaptive, and Markov-driven to capture complex trust dynamics, together with unsupervised trust modeling, real-time RL-based adaptation, cross-domain interoperability, large-scale evaluation, privacy preservation, and quantitative analysis of energy and computational costs for deployment readiness.
}

\bibliographystyle{IEEEtran}
\bibliography{bibfile}

\begin{thebibliography}{10}
\providecommand{\url}[1]{#1}
\csname url@samestyle\endcsname
\providecommand{\newblock}{\relax}
\providecommand{\bibinfo}[2]{#2}
\providecommand{\BIBentrySTDinterwordspacing}{\spaceskip=0pt\relax}
\providecommand{\BIBentryALTinterwordstretchfactor}{4}
\providecommand{\BIBentryALTinterwordspacing}{\spaceskip=\fontdimen2\font plus
\BIBentryALTinterwordstretchfactor\fontdimen3\font minus
  \fontdimen4\font\relax}
\providecommand{\BIBforeignlanguage}[2]{{%
\expandafter\ifx\csname l@#1\endcsname\relax
\typeout{** WARNING: IEEEtran.bst: No hyphenation pattern has been}%
\typeout{** loaded for the language `#1'. Using the pattern for}%
\typeout{** the default language instead.}%
\else
\language=\csname l@#1\endcsname
\fi
#2}}
\providecommand{\BIBdecl}{\relax}
\BIBdecl

\bibitem{10905049}
D.~Saxena and A.~K. Singh, ``A self-healing and fault-tolerant cloud-based
  digital twin processing management model,'' \emph{IEEE Transactions on
  Industrial Informatics}, vol.~21, no.~5, pp. 4233--4242, 2025.

\bibitem{woollacott2025healthcare}
E.~Woollacott, ``Healthcare organizations report rampant email security
  failures – and microsoft {365} is often the weakest link,''
  https://www.itpro.com/business/business-strategy/healthcare-organizations-report-rampant-email-security-failures-and-microsoft-365-is-often-the-weakest-link,
  Sep 2025.

\bibitem{bitsight2025}
\BIBentryALTinterwordspacing
{Bitsight}, ``Inside cyber threats on manufacturing in 2025,'' 2025. [Online].
  Available:
  \url{https://www.bitsight.com/blog/inside-cyber-threats-in-manufacturing-2025}
\BIBentrySTDinterwordspacing

\bibitem{lu2020communication}
Y.~Lu, X.~Huang, K.~Zhang, S.~Maharjan, and Y.~Zhang, ``Communication-efficient
  federated learning for digital twin edge networks in industrial iot,''
  \emph{IEEE Transactions on Industrial Informatics}, vol.~17, no.~8, pp.
  5709--5718, 2020.

\bibitem{tang2025secure}
Y.~Tang, K.~Wang, D.~Niyato, J.~Li, O.~A. Dobre, and T.~Q. Duong, ``Secure data
  sharing and prediction with digital twin and blockchain in healthcare,''
  \emph{IEEE Communications Magazine}, 2025.

\bibitem{soula2024real}
M.~Soula, B.~Mbarek, and A.~Meddeb, ``A real-time trust management model using
  digital twin in iot networks,'' \emph{IEEE Access}, 2024.

\bibitem{10290991}
Z.~Lyu, C.~Cheng, H.~Lv, and H.~Song, ``Blockchain based intelligent resource
  management in distributed digital twins cloud,'' \emph{IEEE Network},
  vol.~38, no.~4, pp. 143--150, 2024.

\bibitem{9830718}
S.~Son, D.~Kwon, J.~Lee, S.~Yu, N.-S. Jho, and Y.~Park, ``On the design of a
  privacy-preserving communication scheme for cloud-based digital twin
  environments using blockchain,'' \emph{IEEE Access}, vol.~10, pp.
  75\,365--75\,375, 2022.

\bibitem{liu2022blockchain}
J.~Liu, L.~Zhang, C.~Li, J.~Bai, H.~Lv, and Z.~Lv, ``Blockchain-based secure
  communication of intelligent transportation digital twins system,''
  \emph{IEEE transactions on intelligent transportation systems}, vol.~23,
  no.~11, pp. 22\,630--22\,640, 2022.

\bibitem{saad2020implementation}
A.~Saad, S.~Faddel, T.~Youssef, and O.~A. Mohammed, ``On the implementation of
  iot-based digital twin for networked microgrids resiliency against cyber
  attacks,'' \emph{IEEE transactions on smart grid}, vol.~11, no.~6, pp.
  5138--5150, 2020.

\bibitem{9112234}
------, ``On the implementation of iot-based digital twin for networked
  microgrids resiliency against cyber attacks,'' \emph{IEEE Transactions on
  Smart Grid}, vol.~11, no.~6, pp. 5138--5150, 2020.

\bibitem{bera2024digital}
B.~Bera, A.~K. Das, and B.~Sikdar, ``Digital twins-empowered secure network
  slice access and isolation for consumer healthcare applications,'' \emph{IEEE
  Transactions on Services Computing}, vol.~17, no.~6, pp. 3429--3444, 2024.

\bibitem{wang2023survey}
Y.~Wang, Z.~Su, S.~Guo, M.~Dai, T.~H. Luan, and Y.~Liu, ``A survey on digital
  twins: Architecture, enabling technologies, security and privacy, and future
  prospects,'' \emph{IEEE Internet of Things Journal}, vol.~10, no.~17, pp.
  14\,965--14\,987, 2023.

\bibitem{mrabet2025towards}
M.~Mrabet and M.~Sliti, ``Towards secure, trustworthy and sustainable edge
  computing for smart cities: innovative strategies and future prospects,''
  \emph{IEEe Access}, 2025.

\bibitem{reiss2011google}
C.~Reiss, J.~Wilkes, and J.~L. Hellerstein, ``Google cluster-usage traces:
  format+ schema,'' \emph{Google Inc., White Paper}, pp. 1--14, 2011.

\end{thebibliography}

\end{document}